\documentclass[journal]{IEEEtran}

\usepackage{setspace}
\usepackage{cite,graphicx,amsmath,amssymb}
\usepackage{subfigure}
\usepackage{fancyhdr}

\usepackage{mdwmath}
\usepackage{mdwtab}
\usepackage{balance}
\usepackage{xcolor}
\usepackage{bm}
\usepackage{amsthm}
\usepackage{threeparttable}
\usepackage{algorithm}
\usepackage{algorithmic}
\usepackage{multirow}
\usepackage{flafter}
\usepackage{makecell}
\usepackage{diagbox}
\usepackage{xcolor}

\begin{document}
\title{Mobile Edge Generation: A New Era to 6G}

\author{\normalsize {
Ruikang~Zhong,~\IEEEmembership{\normalsize Member,~IEEE,}
Xidong~Mu,~\IEEEmembership{\normalsize Member,~IEEE,}
Yimeng~Zhang,~\IEEEmembership{\normalsize Student~Member,~IEEE,}\\
Mona~Jabor,~\IEEEmembership{\normalsize Senior~Member,~IEEE,}
Yuanwei~Liu,~\IEEEmembership{\normalsize Fellow,~IEEE}
}

\thanks{Ruikang~Zhong, Xidong~Mu, Mona Jabor, and Yuanwei~Liu are with the school of School of Electronic Engineering and Computer Science, Queen Mary University of London, London E1 4NS, U.K. (e-mail: r.zhong@qmul.ac.uk; xidong.mu@qmul.ac.uk; m.jabor@qmul.ac.uk; yuanwei.liu@qmul.ac.uk).}

\thanks{Yimeng Zhang is with the State Key Laboratory of Networking and Switching Technology, Beijing University of Posts and Telecommunications, Beijing 100876, China (email: yimengzhang@bupt.edu.cn).}

}

 \markboth{\textit{A Manuscript Submitted to The IEEE Network} }{}

\date{\today}
 \maketitle

\begin{abstract}

A conception of mobile edge generation (MEG) is proposed, where generative artificial intelligence (GAI) models are distributed at edge servers (ESs) and user equipment (UE), enabling joint execution of generation tasks. Various distributed deployment schemes of the GAI model are proposed to alleviate the immense network load and long user queuing times for accessing GAI models. Two MEG frameworks are proposed, namely the single-ES framework and the multi-ESs framework. 1) A one-to-one joint generation framework between an ES and a UE is proposed, including four specific single-ES MEG protocols. These protocols allow distributed GAI models to transmit seeds or sketches for delivering information efficiently. 2) Several protocols are proposed for multi-ESs MEG, which enable multiple ESs to perform the generation task cooperatively or in parallel. Finally, a case study of a text-guided-image-to-image generation is provided, where a latent diffusion model is distributed at an ES and a UE. The simulation results demonstrate that the proposed protocols are able to generate high-quality images at extremely low signal-to-noise ratios. The proposed protocols can significantly reduce the communication overhead compared to the centralized model.

\end{abstract}

\section{Introduction}

Artificial intelligent generated content (AIGC) has shown remarkable success in 2022. Generative artificial intelligence~(GAI) strives to produce fresh content that appears identical or, in some cases, indistinguishable from that of human origin. Models of GAI can produce text, images, audio, and even video \cite{xu2023unleashing}. Noteworthy examples of  GAI include but are not limited to the Generative Pre-trained Transformer (GPT) developed by OpenAI and Wavenet developed by DeepMind \cite{openai2023gpt4}. These models have received widespread application, resulting in substantial network traffic volumes and considerable social impact. It is probable that GAI services will increasingly prevail in future networks \cite{chen2023revolution}.

It is noteworthy that the advent of GAI is likely to be an important trigger for the transition from the fifth generation~(5G) to the sixth generation (6G) of mobile communications \cite{9766098}. Looking back at the history of communication systems, emerging applications have consistently driven the development of networks. From the inception of the analog communication system to the 5G communication system, we have incrementally achieved reliable and efficient transmission of text, voice, images, and video \cite{8114722}. The GAI model has become a new information source, and the generated content needs to be transmitted to users through the network. Given the popularity of GAI, the data delivery for the GAI models has stringent requirements and heavy traffic load on the network. Thus, predictably, ensuring sufficient support for the operation of GAI applications in order to provide users with satisfactory experiences is a key problem for the 6G network.

\subsection{Background of Generative AI}

Research on AIGC was first initiated in the field of computer science for generating different modals of data, including but not limited to text, image, and video \cite{zhang2023complete}. In 2020, the advent of Denoising Diffusion Probalistic Models (DDPM) \cite{ho2020denoising} made an impressive impact on high-quality image synthesis, leading to the development of AIGC based on the diffusion model.  AIGC has also been studied for video generation, where the framework proposed in \cite{guo2023animatediff} can combine the generated static images with motion dynamics to generate temporally smooth animations. Although these GAI methods provide the foundation for obtaining high-quality generated products, they also result in massive complexity \cite{zong2022survey}.


In fact, to solve the problem of high computational complexity caused by the massive parameters of the AIGC model, there have been some studies that have tried to use mobile edge computing (MEC) as a solution to provide some computational power. In \cite{10172151}, the authors proposed an AIGC framework based on cooperative distributed diffusion, which improves the efficiency and edge computing resource utilization of AIGC tasks by leveraging collaboration between devices in wireless networks. The resource constraints of the mobile edge networks for AIGC is considered in \cite{wang2023unified}, and the authors proposed a framework to utilize wireless perception to guide GAI specifically for virtual character generation. However, most of the above-mentioned papers make technical insights and contributions from the perspective of computing power, but the research on the network burden caused by AIGC is still in the initial stage.

\subsection{Motivation of MEG}

Although GAI has powerful functions and practicability, existing centralized GAI still has several problems, including but not limited to causing huge communication overhead, long response time, and lack of diversity.  We can describe the detail problems of the deployment and invocation of existing GAI models as follows:

\begin{itemize}
    \item \textbf{Huge communication overhead:} It is foreseeable that the centralized AIGC solution will have a high communication overhead since the user's generation request and the generated content need to be uploaded and downloaded over the network. Existing commercial generation models, such as GPT series production developed by Open AI, mainly support the input and output of text and images. In the future, once high-definition video becomes the mainstream product, the amount of data stream required by GAI will become a heavy tax on the network resources. Even if the MEC server is called to supplement computing resources, the data-delivering overhead between the edge server (ES) and the user device will not be reduced. Therefore, how to solve the huge network load caused by generative AI is a new challenge in the communication and network society.

    \item \textbf{Response timeliness:} The overall response time on the user side is another major issue for GAI service. Firstly, the existing GAI service is prone to congestion and even paralysis due to traffic overload \cite{GPTqueue}. The long queuing time to obtain the AI service caused damage to the response timeliness, thus, a distributed solution is required. It is worth noting that the overall response time does not refer to the time when the GAI server starts responding to the generation task, but ought to be the time from when the user submits the application to receiving satisfactory generated content. In the existing GAI framework, only one central model is providing service. Therefore, the user may need to repeat the request multiple times until they receive satisfying results, which increases the overall response time and further exacerbates the \textbf{communication overhead problem}.

    \item \textbf{Diversity and personalization:} Although the GAI model is well-trained, the diversity of products produced by a single model is difficult to ensure \cite{coman2011generating}. Users from different countries, locations, or ethnicities are likely to have different preferences. The centralized model may not able to capture the preference of the users at the first time, or even the centralized model may not be able to produce qualified generated content for some specific generation requests. Using MEG enables edge servers to train their model by customized data sets. As a result, GAI models of different edge servers tend to generate preferred content. Even some models deployed on the UE can understand the specific preferences of each user and achieve customized generation.
\end{itemize}

In order to resolve the above mentioned problems, this article is dedicated to designing a closely integrated GAI and network model, named mobile edge generation~(MEG).


\subsection{Contributions}

In an effort to address the aforementioned problems, we propose the framework of MEG in this paper, which deploys the GAI model in a distributed manner at the ESs and UEs. Beyond MEC, this framework integrates the distributed model deployment, distributed computation, joint training, and generation. We studied the joint generation between a single ES and a single UE, as well as the joint generation between multiple ESs and a single UE. The proposed MEG framework generates seeds or sketches to transfer information between multiple distributed model components, which considerably reduces the communication resource consumption of AIGC. We tested the proposed MEG protocols in a text-guided image-to-image generation. The simulation results demonstrate that with an extremely low signal-to-noise ratio of -20dB, the proposed protocol can generate quality images having minor distortion pixels, and the proposed MEG protocol can significantly reduce the communication overhead.

\section{Single-ES MEG Model}

In this section we propose a MEG network framework that enables an ES and an UE to jointly operate contents generation.

\begin{figure}[htb]
\centering
\includegraphics[width=0.5\textwidth]{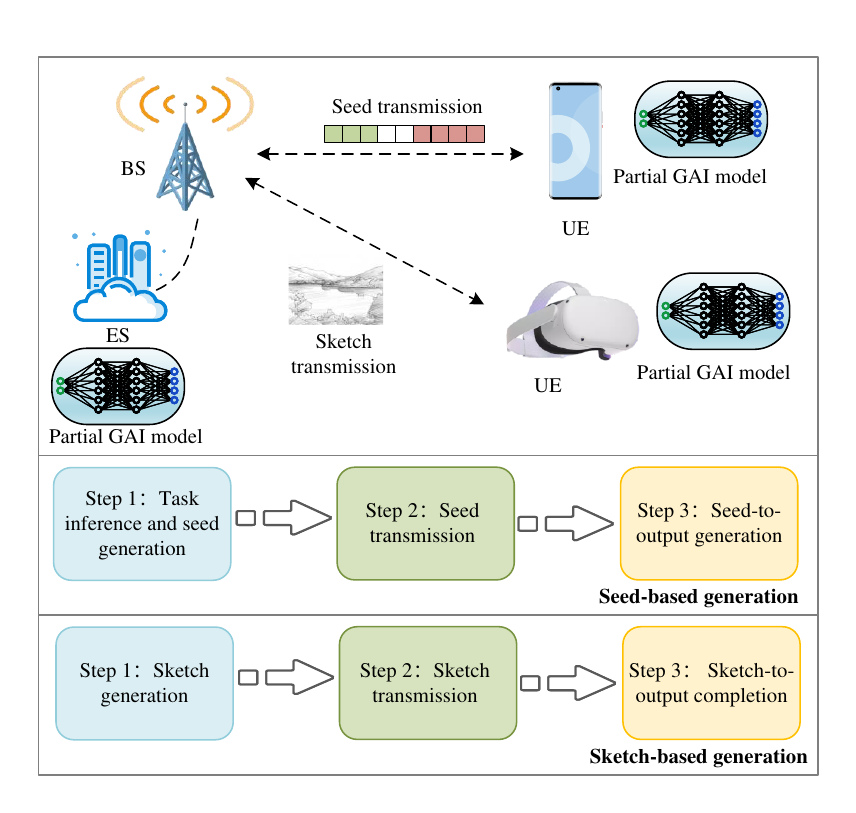}
\caption{The working flow of the seed-based and the sketch-based MEG.}
\label{Fig.1}
\end{figure}

\subsection{Network Model}

\begin{figure*}[t!]
\centering
\includegraphics[width=0.9\textwidth]{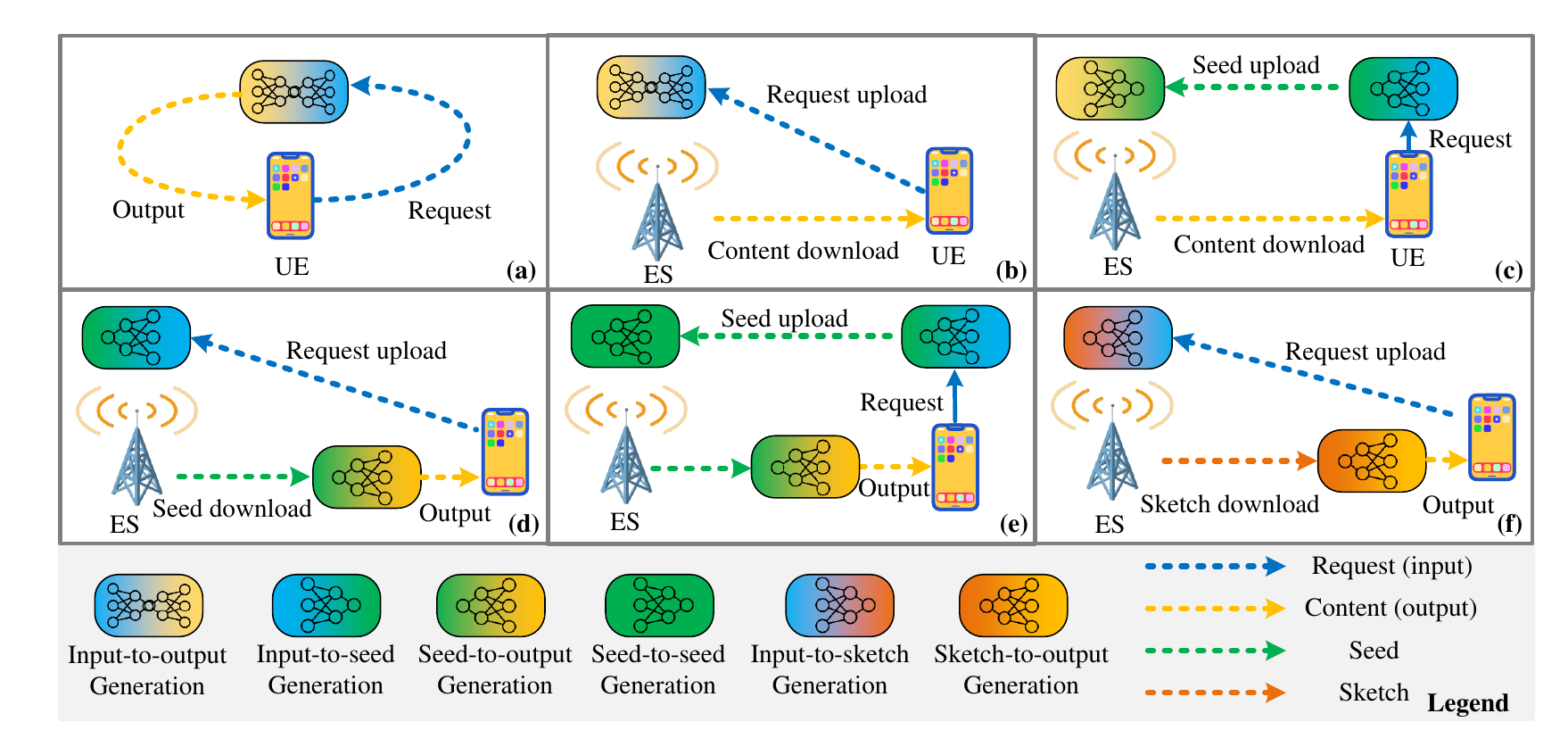}
\caption{MEG Protocols. (a) Local generation; (b) Centralized generation; (c) UIBG protocol; (d) BIUG protocol; (e) CIAG protocol; (f) ESUC protocol.}
\label{Fig.2}
\end{figure*}

We consider a two-layer network model consisting of ES and UEs as displayed in Fig. \ref{Fig.1}. We assume that UEs, also known as edge devices, such as cell phones, personal computers, etc., are operated by users, receiving generation requests from users. The form of generating requests may be diverse, including but not limited to image-to-image generation, text-to-image generation, and multi-modal data generation, etc. Users can exchange data with edge servers deployed in base stations via wireless channels. We assume that the ESs are connected to the center/cloud server through optical fiber, and the ES has stronger computing and storage capabilities than UEs. In fact, the data acquisition, model update, and other operations of ESs rely on the data and computing power of the cloud server. Therefore, a three-layer model of cloud-ESs-UEs is desirable in practice. Since this article focuses on the generation of edge networks, in order to concisely present the proposed network model, we neglect the operations and overhead of edge servers accessing cloud servers.

Existing GAI schemes generally deploy the model on one hardware, whether it is on the user device or on the central server. These two schemes are depicted in Fig. \ref{Fig.2}(a) and Fig. \ref{Fig.2}(b). These two solutions have similar operation processes, which is, the GAI model takes the user request as input and generates content (output) directly according to the request from the user. We call this approach request-based generation. Different from these conventional solutions, the core idea of single-ES MEG is to split the GAI model into multiple modules and deploy them on ESs and UEs, respectively, thereby making full use of the computing power of all devices\footnote{It is worth noting that the proposed MEG framework does not specify any specific GAI model, such as a GAI model based on GPT or diffusion. Different GAI models can have different detailed designs on distributed deployment and information exchange under the principle of MEG.}.

\subsection{Communication Schemes for MEG}

As shown in Fig. \ref{Fig.1}, the operating process of MEG is divided into three basic steps, namely 'inference-transmission-generation'.

\subsubsection{Seed-based Generation Scheme}

The first step is seed creation, where the partial of the GAI model 'inferencer' is responsible for extracting key semantic features from the information provided by the user request. We name the extracted features as the 'seed' for the generation task. The second step is to send this seed to the second part of the distributed generative artificial intelligence (DGAI) model. The second part of the DGAI model 'generator' is deployed on another device. Since the size of the seed is generally significantly smaller than the original input from the user, this will save considerable communication resources. Finally, the second part of the DGAI model will generate the content requested by the user based on the seed. We name this scheme seed-based generation.

\subsubsection{Sketch-based Generation Scheme}

Different from the seed-based generation mode, the sketch-based generation mode concentrates the core work of generation on a single device. The two parts of DGAI models in this scheme play the roles of 'sketcher' and 'completer', respectively. The sketcher creates a basic outline based on user input, which means that the inference and generation work is done by the sketcher. However, the sketches produced by the sketcher may have extremely low resolution and even lack some details. After the sketch is generated, it is sent to DGAI's second model, the completer. The completer enhances and creates secondary works based on the sketch.

\subsection{Operating Protocols}

Both the seed-based MEG scheme and the sketch-based MEG scheme divide the generation task into two subtasks. What needs to be determined next is which type of device to execute the subtasks on. Specifically, we propose four following task allocation protocols.

\subsubsection{\textbf{UE inference and Edge generation (UIEG)}}

In the first MEG protocol, the inferencer is deployed at the UE, the user request is first parsed at the UE, and seeds are generated for the generation task. Next, the seed is transmitted to the ES, and the generator deployed at the edge server completes the generation based on the information carried by the seed. Finally, the generated content needs to be sent to the UE. The advantage of this solution is that it can effectively save uplink bandwidth since only the seed but not user input needs to be uploaded to the server. As a result, this protocol is clearly suitable for generation tasks with complex inputs, such as generating pictures from pictures, or text from videos, etc.

\subsubsection{\textbf{Edge inference and UE generation (EIUG)}}

The opposite of the UIEG protocol is the EIUG protocol. As the name suggests, the EIUG protocol deploys inferencer at the ES, and the generator is deployed at the UE side. In this protocol, the complete information of the user request needs to be transmitted to the ES, and then the ES returns the seed to the UE to complete the generation. Contrary to UIEG, EIUG can significantly reduce downlink bandwidth consumption. This protocol is suitable for video-to-text, video-to-speech generation, e.g. AI-generated dubbing.

\begin{figure*}[htb]
\centering
\includegraphics[width=0.9\textwidth]{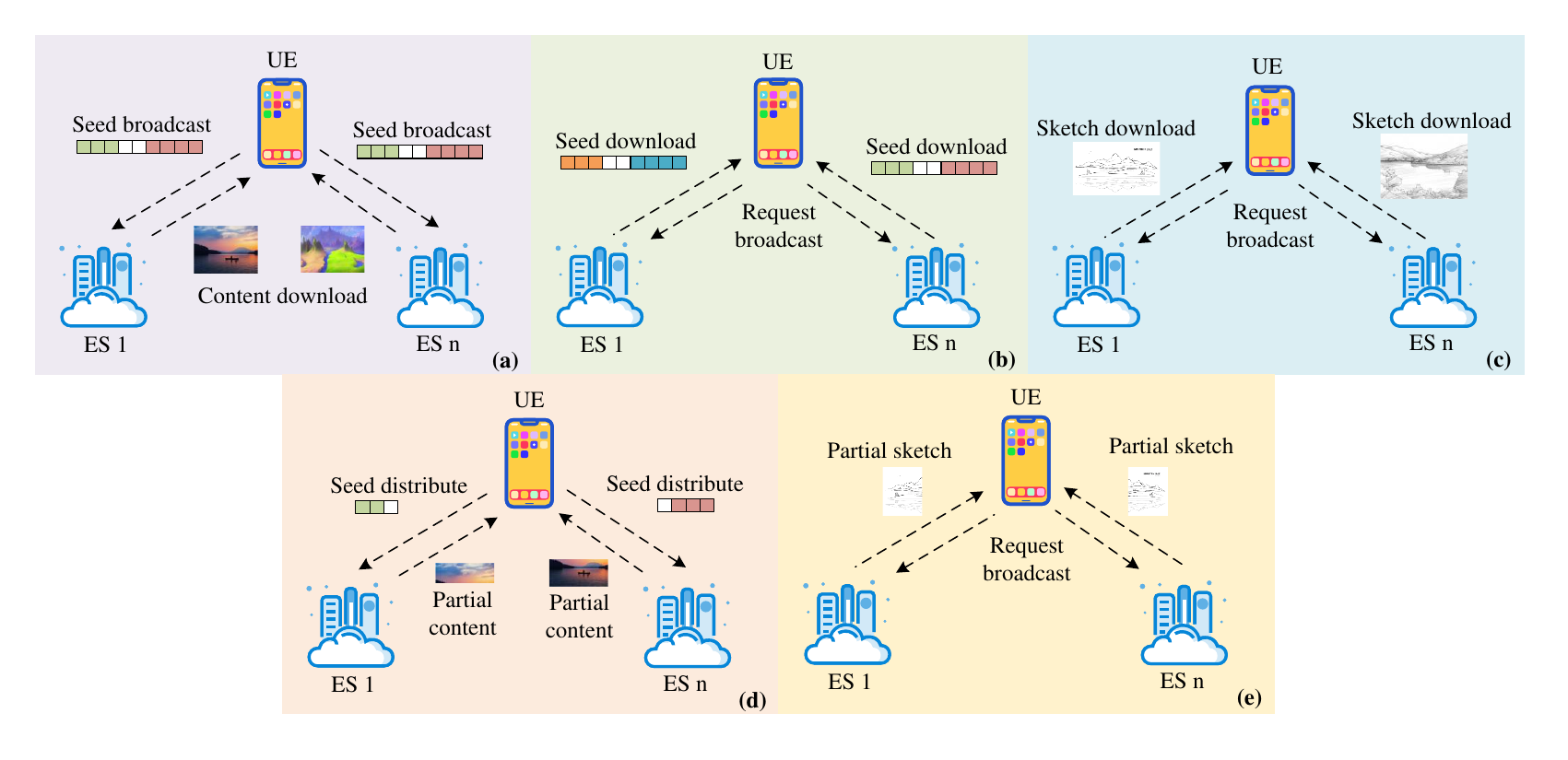}
\caption{Multi-ES MEG Model (a) UIDG protocol; (b) DIUG protocol; (c) DSUC protocol; (d) UIDCG protocol; (e) DCSUC protocol.}
\label{Fig.3}
\end{figure*}

\subsubsection{\textbf{Cooperative inference and Generation (CIAG)}}

The CIAG protocol completely gets rid of transmitting the original input or output from users or GAI models. The input from the users is analyzed at the UE first, and a seed containing the feature of the task is generated. After that, the seed of the task has to be sent to the ES. The ES will perform a seed-to-seed generation, which transfers the seed of the task to a seed of the generated content. Finally, the seed of content needs to be sent back to the UE to generate human-readable content. It can be considered that the content has been generated at the ES, but it is represented in the form of features rather than the final form for presentation. This method can save communication bandwidth resources to the maximum extent since the consumption of transmitting seed is much less than any input and output content.
Compared with the UIEG protocol, the EIUG and CIAG protocols only need to transmit seeds rather than the generated content. Since seeds have better noise resistance, it is more likely to present quality-generated products to users.

\subsubsection{\textbf{Edge sketch and UE Complete (ESUC)}}

For the sketch-based MEG scheme, we propose the ESUE protocol. Initially, the UE forwards the user requirements to the ES, and then the ES is responsible for generating the sketch. The UE subsequently completes the sketch according to its specific requirements by applying techniques such as simple AI resolution enhancement or regeneration based on the sketch. The transmission of sketches involves a slightly higher communication overhead compared to transmitting seeds. However, this protocol has the advantage of asymmetrically distributing its data requirements and computational complexity between the two devices. Consequently, ESs are utilized to produce sketches as they have easier access to superior models and greater computing power. In addition, due to the above reason, we have no intention of utilizing UE as a sketcher.

\section{Multi-ES MEG}

In order to further improve the response speed of GAI services, multiple ES can be jointly employed for a generation task. Thus, we propose many-to-one generation frameworks that enables multi-ES to serve a UE in parallel or cooperatively.

\subsection{Network Model}

We expand the single-ES MEG model described in section 2 to a case that a UE is associated with multiple edge servers as shown in Fig. \ref{Fig.3}. One notable benefit of the many-to-one joint generation approach is the capacity to distribute the computational complexity across various ESs. Furthermore, since each ES may operate a distinct model, this approach has the potential to enhance the diversity of the content produced by GAI, facilitating users' access to the desired content and circumventing repetitive regeneration.

There are two ways to achieve multi-ES joint generations. The first one is parallel generation, where the user release the generation task to multiple ESs, and each ES completes the generation task independently and in parallel. The other approach is

\subsection{Parallel Protocols for Multi-ES MEG}

For the multi-ES GAI services, we propose the following protocols that each ES working independently to complete the generation task in parallel:

\subsubsection{\textbf{UE inference Distributed generation (UIDG)}}

The UIDG protocol can be regarded as an extension of the UIEG protocol. As presented in Fig. \ref{Fig.3}(a), the UE is responsible for the seed generation. The seeds are then distributed to multiple edge servers. The delivery of the seeds can be broadcast through a wireless channel, or forwarded by a relay server. Each edge server will implement seed-based generation in parallel, and then send the generated content to the UE. The advantage of the UIDG protocol is that multiple edge servers can generate using different models, enhancing the chances of users obtaining the desired generation promptly. However, the disadvantage of this solution is that the edge server needs to transmit the generated content to the UE in sequence, which may require high coordination capabilities among the edge servers. In addition, transmitting the generated content multiple times may cause considerable communication overhead.

\subsubsection{\textbf{Distributed inference UE generation MEG (DIUG)}}
As presented in Fig. \ref{Fig.3}(b), in the DIUG protocol, task objectives are disseminated by the UE to several edge servers. Afterward, the seed is produced and conveyed back to the UE by the server. Using the seed as a basis, the UE can generate content for the user. This protocol diminishes communication overhead substantially. However, the UE is obliged to produce various seeds individually. This necessitates the UE to possess relatively strong computational capabilities.

\subsubsection{\textbf{Distributed sketching and UE Complete (DSUC)}}

DSUC is a sketch-based generation scheme as shown in Fig. \ref{Fig.3}(c). Similar to the DIUG, task requests need to be forwarded to edge servers. The edge servers are then responsible for generating the sketches independently, using the same theory described in the ESUC protocol. This protocol resembles DIUG, yet its strength lies in the sketch's human readability. The UE does not have to produce appropriate content for every sketch anymore. Users can review the sketch first and pick the one that they are satisfied with. Subsequently, the UE generates content based on the selected sketch. This procedure balances the UE's computational complexity and communication overhead.

\subsection{Cooperative Protocols for Multi-ES MEG}

The parallel protocol enables several edge servers to deliver various output contents to users within a single generation time. Although such parallel approaches are efficient, they cannot shorten the time of single generation task. To reduce the response time of a single generation task, we designed cooperative generation protocols for seed-based generation and sketch-based generation, respectively.

\subsubsection{\textbf{UE inference Distributed Cooperative generation (UIDCG)}}

The same as the UIDG protocol, the UE generates a seed. However, as illustrated in Fig. \ref{Fig.3}(d) it is no longer broadcasted but divided into several sub-seeds and dispersed among multiple edge servers. Each edge server generates partial content based on the sub-seed it received. Then the components of these contents will be sent to the UE separately, and the UE is responsible for merging these components into the final output. The cornerstone of accomplishing distributed cooperative generation is the joint training of multiple components of the model. Models deployed at different edge servers and the UE need to be trained jointly by a loss function. The significant advantage of UIDCG protocol is that it enables multiple servers to work on the same generation task, but at the expense of joint training of multiple servers, which places higher requirements and expenses on model management.

\subsubsection{\textbf{Distributed Cooperative sketching and UE Complete (DCSUC)}}

In the DCSUC protocol, the UE needs to forward input from the user to multiple edge servers. However, edge servers are jointly trained to generate only partial sketches. Taking the image generation for mountains in Fig. \ref{Fig.3}(e) as an example, each server generated a sketch of part of the mountain range. The two parts will be sent to the user to be merged into a complete sketch. Finally, the UE is regenerated based on the sketch. The DCSUC protocol enables each edge server to set up a model with a smaller scale, effectively reducing the waiting time for generation. However, since multiple servers are jointly trained, how to maintain their respective abilities to independently complete the generation task is a problem.

\begin{figure*}[t!]
\centering
\includegraphics[width=0.7\textwidth]{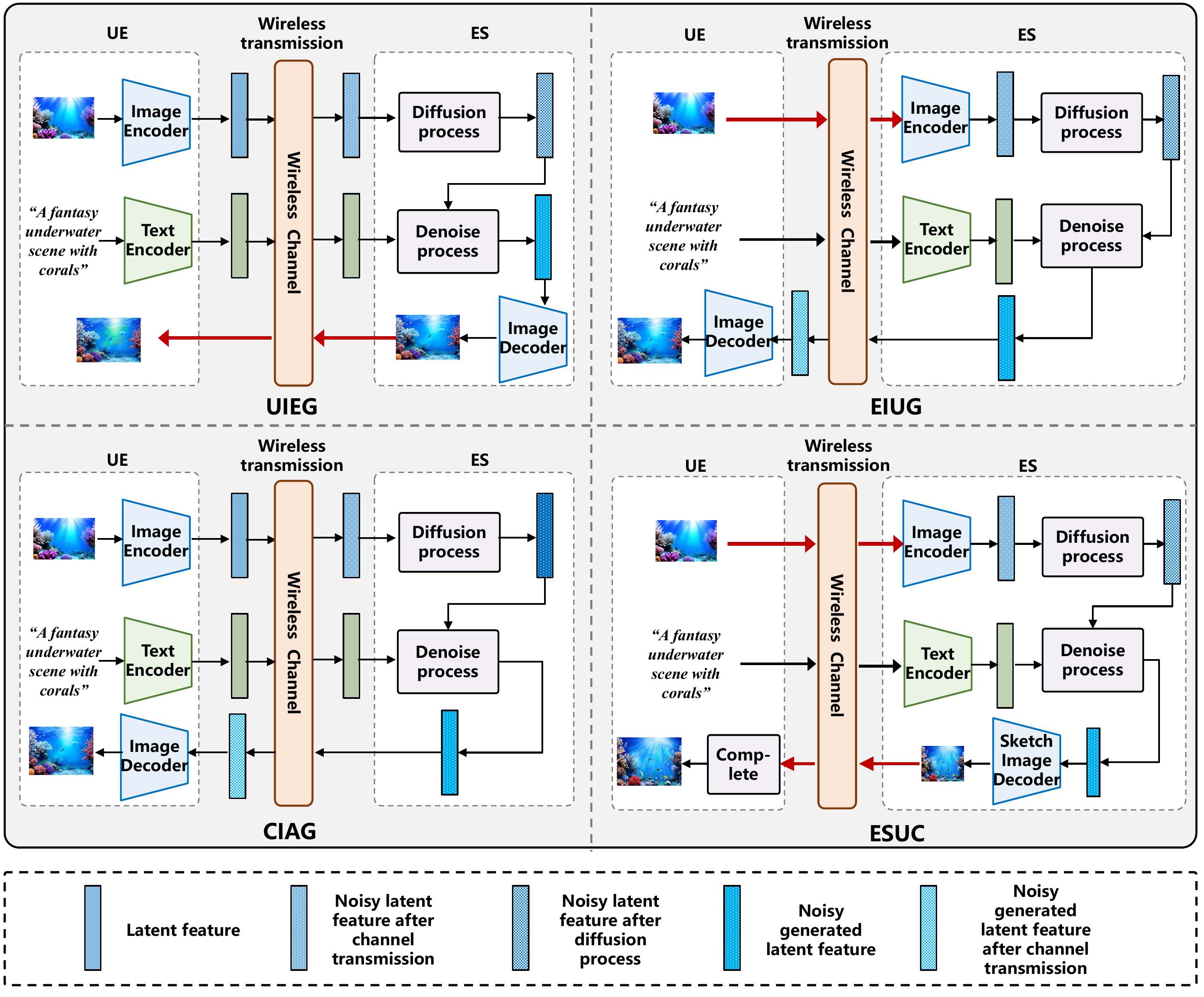}
\caption{Work flow of a distributed LDM model in MEG.}
\label{Fig.4}
\end{figure*}

\section{Case study}

To verify the performance of the proposed MEG protocols, in this section, we provide a case study of a text-guided-image-to-image generation. We perform computer simulations for all MEG protocols presented in Section II. A Latent Diffusion Model (LDM) proposed in \cite{Rombach_2022_CVPR} is invoked as the backbone of the proposed MEG model. Specifically, given a text prompt describing the intended images, LDM can generate diverse images with a similar style to an input image. The generation speed is highly improved by mapping the original images with an image encoder to the low-dimension latent features, based on which the diffusion process is proceeded to obtain the noisy latent features. By denoising the noisy latent features, the generated latent features are produced under the conditioning of the text prompt, which is also embedded as text representations by a text encoder. Finally, the intended images are reconstructed with an image decoder based on the generated latent features.

\begin{figure*}[t!]
\centering
\includegraphics[width=0.8\textwidth]{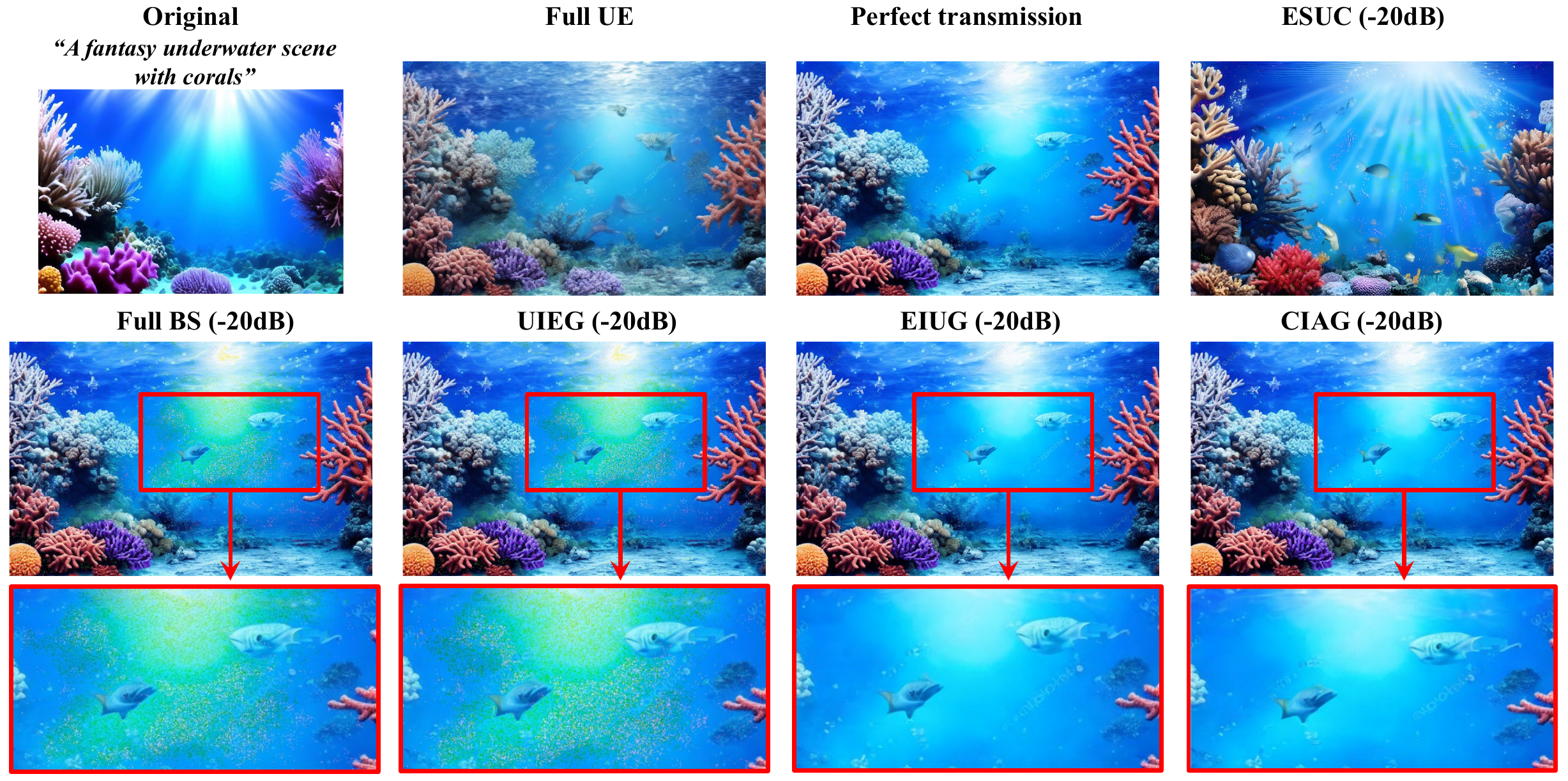}
\caption{Visible results text-guided image-to-image generation.}
\label{Fig.5}
\end{figure*}

Based on the well-trained weights of LDM, we implemented four single-ES MEG protocols. The channel between the UE and ES is considered an Additive White Gaussian Noise (AWGN) channel. The specific pipelines are illustrated in~\ref{Fig.4}. For UIEG, the image and text encoders are deployed at the UE to represent the original images as the latent features (seed) for uplink wireless transmission. At the ES side, based on the contaminated latent features, the generated latent features are also obtained with wireless noise included. Then the intended images of the UE are recovered with the image decoder and transmitted back to the UE. Opposite from UIEG, BIUG directly transmits original images during the uplink stage while feedback on the noisy generated latent features to the UE, where the image decoder is deployed for image reconstruction. Combining the advantages of the above two protocols, CIAG implemented the image representation and reconstruction both at the UE, while the transmission costs were reduced by only exchanging the latent features with the ES. ESUC follows the same uplink transmission process as BIUG, while a sketch image decoder is alternatively deployed at the ES to generate sketches of the intended images. Then the sketch images are transmitted to the UE, where a complete module is adopted to transform the sketch images into the intended images. Note that the text transmission of four protocols is all assumed to be lossless.

\begin{figure*}[t!]
\centering
\includegraphics[width=0.95\textwidth]{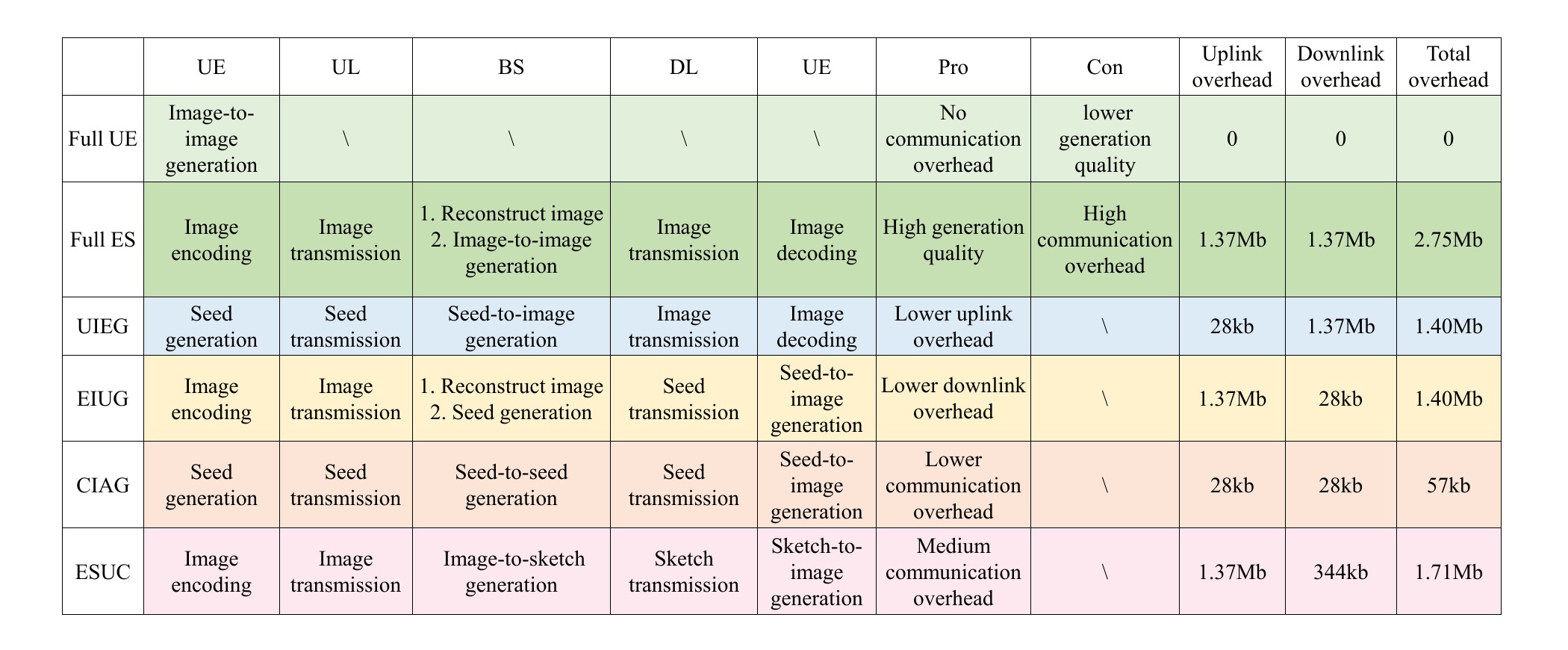}
\caption{The communication overhead of each MEG protocol.}
\label{Fig.6}
\end{figure*}

Fig. \ref{Fig.5} presents the visible text-guided image-to-image generation results. We tested the protocols under -20dB AWGN channel, and the results are compared to the baseline with perfect transmission. The text command instructs the MEG model to generate corals in an underwater scene. All of the proposed protocols are able to generate qualified images following the text instructions. Although the \textbf{full UE} generation does not suffer from any distortion caused by the transmission, it has incorrect image details such as twisted fishes due to the limited local computation power. Some subtle distorted points can be observed in the \textbf{full BS} and \textbf{EIUG} protocol due to the noise during the image download. The \textbf{EIUG} and \textbf{CIAG} protocols generated images having perfect details since the transmission of features is more robust as analyzed in the subsection of \textbf{CIAG}. Finally, the image generated by \textbf{ESUC} protocol shows better light and shadow effects, but the fishes lack some details.

Fig. \ref{Fig.6} provides the quantitative analysis of the communication overhead of the proposed MEG protocols. Firstly, the specific operation needed for each protocol is specified in Fig. \ref{Fig.6}. In this simulation, the user input consists of an image and a text command, and in fact, they need to be transmitted to achieve text-guided image-to-image generation. Due to limited space in the table, we simplify the description to image-to-image generation. It can be observed that directly transmitting images not only requires a larger bandwidth but also creates the complexity of encoding and decoding the images. Regardless of whether a centralized AIGC or MEC-assisted AIGC framework is employed, the communication overhead is difficult to be significantly reduced.

Having a more specific compression, the two benchmark solutions, which are completely generated in the UE or ES, will cause a huge computing load to the UE or a huge communication overhead to the network. The \textbf{UIEG} and \textbf{EIUG} protocols reduce the uplink and downlink transmission consumption from 1.3Mb to 28kb respectively. As outlined in Section 2, the \textbf{CIAG} scheme has minimal bandwidth consumption since only latent features are transmitted, consuming only 57kb of transmission bandwidth. This approach decreases the aggregate volume of transmitted data by a factor of 47.2 while maintaining the excellence of the produced precursors. Finally, the communication overhead of the \textbf{ESUC} protocol is significantly lower than the centralized scheme, but higher than all seed-based generation protocols. However, it is worth noting that the ESUC protocol always retains visual graphics, allowing users to preview thumbnails of upcoming content.

\section{Conclusions \& Open Problems}

\subsection{Concluding Remarks}
We proposed a MEG framework to closely integrate the AIGC and the edge network, distributing the computation and storage resources required by the AIGC model to edge servers,  significantly reducing the communication overhead for the cooperation of the ESs and UEs. We investigated the one-to-one single-ES MEG framework and the many-to-one multi-ES MEG frameworks. In the case study, we verified the performance in terms of generation quality and communication overhead of the proposed MEG protocols for text-guided image-to-image generation tasks. The simulation results demonstrated the proposed protocols can successfully work in cruel channel conditions, having sufficient low communication overhead.

\subsection{Open Problems}

\subsubsection{Management of distributed GAI models}
Different from edge caching and edge computing, model management is a unique problem in MEG. Proper model management is crucial in MEG as it determines the quality of the generated products. However, model management is highly challenging due to the enormous storage space required by the GAI model and the potential pitfalls of model updates. Thus, the storage of distributed GAI models has to jointly consider both the storage and computing capability of the devices. In terms of model updates, federated learning could potentially offer a resolution for swift updates and online training of models across multiple ES. However, balancing the personalization and accuracy of the model poses a significant challenge.

\subsubsection{Data security and privacy}

As the MEG solution necessitates data transmission between multiple devices, possible data security concerns emerge in the communication progress. During the MEG task, the seeds or sketches contain a considerable quantity of data. If they were to be intercepted by an eavesdropper, the eavesdropper could perform the generation tasks illegally and without authorization based on the eavesdropped seeds, resulting in a substantial risk to the users' privacy. Therefore, how to ensure that seeds are not eavesdropped or forged during ES and transmission is a key issue.

\subsubsection{Resource management}

Completing the MEG tasks requires jointly calling various types of resources, such as the access of distributed GAI models, data catch, computing resources, and communication resources. These resources are not allocated and operated independently. For example, the availability of the model and computing resources have to be guaranteed simultaneously on the same device; collaborative generation tasks of multiple ES require distributed resource management. Therefore, joint management of multiple resources will be a critical challenge for improving the efficiency of MEG.


\bibliography{Network_ref_abb}
\bibliographystyle{IEEEtran}

\end{document}